# FRAMEWORK FOR WIRELESS NETWORK SECURITY USING QUANTUM CRYPTOGRAPHY

Priyanka Bhatia[1] and Ronak Sumbaly[2]

[1,2] Department of Computer Science, BITS Pilani Dubai, United Arab Emirates

### ABSTRACT

*Data that is transient over an unsecured wireless network is always susceptible to being intercepted by anyone within the range of the wireless signal. Hence providing secure communication to keep the user's information and devices safe when connected wirelessly has become one of the major concerns. Quantum cryptography provides a solution towards absolute communication security over the network by encoding information as polarized photons, which can be sent through the air. This paper explores on the aspect of application of quantum cryptography in wireless networks. In this paper we present a methodology for integrating quantum cryptography and security of IEEE 802.11 wireless networks in terms of distribution of the encryption keys.*

### KEYWORDS

*IEEE 802.11, Key Distribution, Network Security, Quantum Cryptography, Wireless Network.*

## 1. INTRODUCTION

Wireless Networks are becoming increasing widespread, and majority of coffee shops, airports, hotels, offices, universities and other public places are making use of these for a means of communication today. They're convenient due to their portability and high-speed information exchange in homes, offices and enterprises. The advancement in modern wireless technology has led users to switch to a wireless network as compared to a LAN based network connected using Ethernet cables. Unfortunately, they often aren't secure and security issues have become a great concern. If someone is connected to a Wi-Fi network, and sends information through websites or mobile apps, it is easy for someone else to access this information.

Wireless networks generate large amount of data, which often is sensitive and vulnerable to interceptions than wired networks. This has increased the risk for users significantly and to combat this consideration, wireless networks users may choose to utilize various encryption methodologies. Encryption is the key to keep information secure online in a Wi-Fi network. When information is encrypted, it's scrambled into a code so others can't get it. Thus, due to the high probability of information compromise associated with Wi-Fi networks, various encryption methods have been developed. However, commonly utilized encryption methods are known to have weaknesses and are susceptible to attackers thereby compromising confidentiality.

In order to make secure communications around a wireless network, communication between nodes (users) and base station (BS) to other nodes should be handled carefully by means of an efficient key management protocol. Quantum Key Distribution (QKD) using quantum cryptography is a new method in key distribution scheme, which allows broadcast of a network key with absolute confidentiality. This method solves the problem of key distribution by the properties of quantum information and provides a secure communication network between two users with unconditional security. This paper presents a methodology for key distribution in a wireless networks using quantum cryptography and its protocols. The paper also discusses the various other methods to keep wireless networks safe and secure.

*Structure of the Paper*

The remainder of the paper is organized as follows: In Section 2, the objective of this paper is presented. In Section 3, the various acronyms used in this paper have been given. Related works are addressed in Section 4. The fundamentals of wireless networks along with its standard protocols and security issues are presented in Section 5. Section 6 deals with the overview of classical cryptography explicitly expressing its methodology and limitations. Quantum Cryptography along with its protocols for communication is discussed in Section 7. A Quantum Key Distribution (QKD) based Wireless Network model for a secure communication system is presented in Section 8. Advantages of using QKD in Wireless Networks are discussed in Section 9. Section 10 presents the specific contribution of the research. Alternative approaches for a secure wireless communication system are presented in Section 11. The paper is concluded in Section 12.

## 2. OBJECTIVE

The present work is intended to meet the following objectives:

1. Investigate potential susceptibility and security issues present in wireless network communication.
2. Provide a framework for unconditional security in wireless networking using quantum key distribution (QKD).
3. Discuss the benefits of Quantum Cryptography in ubiquitous computing.

## 3. ACRONYMS

AES     : Advanced Encryption Standards
AP      : Access Point.
ARP     : Address Resolution Protocol.
BS      : Base Station.
DoS     : Denial of Service.
EAP     : Extensible Authentication Protocol.
EAPOL: Extensible Authentication Protocol over LAN.
GTK     : Group Temporal Key.
IEEE    : Institute of Electrical and Electronics Engineers.
KCK     : Key Confirmation Key.
KEK     : Key Encryption Key.
LAN     : Local Area Connection.
MAC     : Media Access Control.
MIC     : Message Integrity Code.
PMK     : Pairwise Master Key
PTK     : Pairwise Transient Key
Qubit   : Quantum Bit.
QKD     : Quantum Key Distribution.
Q-MIC : Quantum Message Integrity Code.
RSNA  : Robust Security Network Association.
TK       : Temporal Key.
TSN     : Transaction Security Network.
WEP    : Wired Equivalent Privacy.
Wi-Fi   : Wireless Fidelity.
WLAN : Wireless Local Area Network.
WPAN : Wireless Personal Area Network.
WPA    : Wi-Fi Protected Access.

## 4. RELATED WORKS

Quantum cryptography is described as a point-to-point secure key generation technology that has emerged in recent times in providing absolute security. Researchers have started studying new innovative approaches to exploit the security of QKD for a large-scale communication system. A number of approaches and models for utilization of QKD for secure communication have been developed.

The uncertainty principle in quantum mechanics created a new paradigm for QKD [21]. One of the approaches for use of QKD involved network fashioned security. BBN DARPA quantum network is an example of such network. Researchers at Boston, Harvard University, and BBN technologies jointly developed the DARPA Quantum Network in 2004 [14]. The main goal was point-to-point Quantum network that exploited QKD technology for end-to-end network security via high speed QKD.

Other approaches and models equipped with QKD in network fashion are introduced in the literature as [15-16]. A different approach that this paper deals with is using QKD in existing protocols, which are widely used on the Internet to enhance security with main objective of unconditional security. Papers [17-20][24] present models and schemes to integrate QKD in classical security protocols like IPsec, PPP and TLS.

Related to the same approach we argue our work presents a new scheme to show how to integrate QKD in wireless network protocols.

## 5. OVERVIEW OF WIRELESS NETWORKS

### 5.1. Wireless Networks

In today's era, everyone wants their necessary data to be handy, portable and accessible from almost every place they visit throughout the day and this is made possible by using wireless networks. Wireless networks [1], as the name suggests, are those networks that are not connected by any physical means such as Ethernet cables and thus provide the user with great mobility and convenience. Also, it saves one from the expenses on the cables that would be required if wired network is chosen as well as makes it easier for moving the base of the devices from location to another by just moving the machine along with the wireless network card.

A wired network helps in point to point transfer, that is, sends data between any two devices that are connected with each other through an Ethernet cable but in case of wireless networks, the transfer of data is a broadcast service where the data is sent to all possible directions in the medium within a limited range as the medium of data transfer is air here and not cables. Wireless networks consist of four basic components: Transmission of data using air waves, access points (AP) to establish a connection to the public or private (organization) network and the wireless client operated by the user. Fig. 1. Shows the basic wireless networking components.

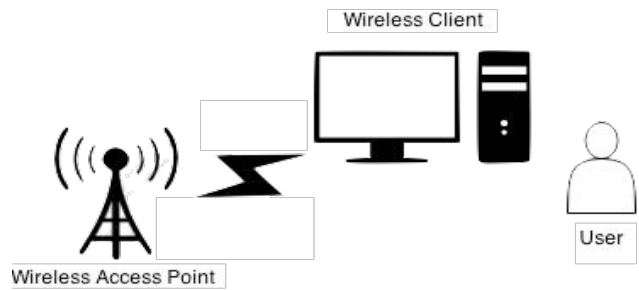

Figure 1. Wireless Networking Components

## 5.2. Security Issues in Wireless Networks

Wireless networks do not promise quality of service during transmission and chances of intrusion into such networks are very high since the transmission here takes place through the medium of air and not cables. So, it doesn't only require protection against uninvited users from accessing the network but also needs to secure the users' private data that is being transmitted. The general security issues for wireless networks are as follows [2]:

1) *Confidentiality*: The data being sent across the network is encrypted during transit so as to ensure that the information is read only by the intended user and hence authentication of the receiver is required as well who will be given the key for the decryption of the received data.

2) *Integrity*: Wireless networks are exposed to attacks that would harm the data's integrity. The integrity prevention methods applied are similar to the ones used in wired networks.

3) *Availability*: Wireless networks are vulnerable to denial of service attacks. Radio jamming can be used to restrict the availability of network. Another attack called, battery exhaustion attack is also arisen where unauthorized users repeatedly send messages or data to connected devices and hence runs down the device's battery.

4) *Eavesdropping and Authentication*: Wireless networks are broadcast as mentioned earlier hence there are more number of access points and these access points can be used to enter the network. Preventing this eavesdropping is necessary.

5) *Blue Snarfing or Blue jacking*: These are the attacks made using Bluetooth in order to tamper data or data theft.

6) *War Driver*: Another type of security attack where a wireless device (like laptops) somehow tries to connect to unprotected network and could record the private data of the user connected to the same network.

## 5.3. Protocols in Wireless Networks

There are wireless network protocols developed in order to provide privacy protection of the user data by encrypting the data being sent across the network. WLANs are defined under the IEEE 802.11 standard [3]. Table 1 summarizes the three major wireless security protocols.

Table 1. Wireless Security Protocols

| Protocol | Full Form | Description |
|---|---|---|
| WEP | Wired Equivalent Privacy | Security similar to wired networks. Provides 10 to 26 long key. Easily broken security algorithm with numerous flaws. |
| WPA | Wi-Fi Protected Access | Uses Pre-shared Key (PSK) and temporal key integrity protocol [4] for encryption of data. |
| WPA2 | Wi-Fi Protected Access V.2 | Uses Advanced Encryption Standards (AES) [5] for encrypting data. |

The security of a WLAN depends on the secrecy of the entire encrypting and decrypting process. Various algorithms are presently used for encrypting and decrypting without bargaining the security of the data being sent. The design and analysis of various mathematical techniques for encryption/decryption of data that ensure secure communications is termed as cryptography.

## 6. CLASSICAL CRYPTOGRAPHY

### 6.1. Overview of Classical Cryptography

Cryptology is defined as the practice and study of techniques for secure communications in presence of adversaries (third parties), which underpin cryptography and cryptanalysis. Fig. 2. Shows the hierarchy between data security and cryptology.

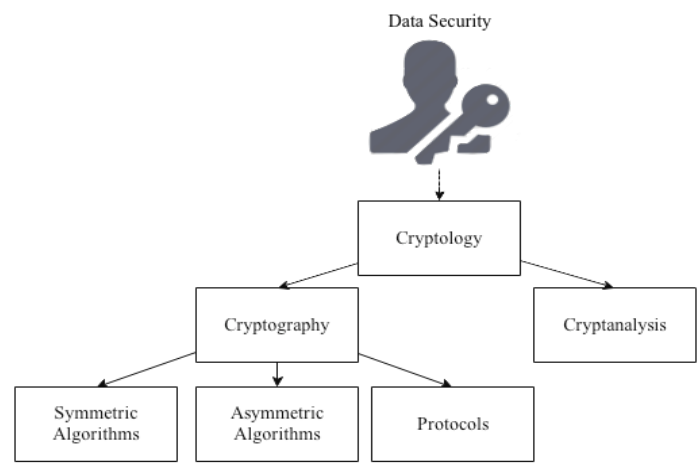

Figure 2. Relationship between Data Security and Cryptology

Cryptography is composed of two major goals:
- *Information privacy:* For keeping information transmitted via a network private without giving any information to a third party and
- *Authentication:* To check the integrity of the message received by the user from another party connected via the same network.

Cryptography operates by a senders' transmittal of an encrypted message in a systematic manner that conceals its original meaning that is then recovered by the receiver who decrypts the transmission to get the original message. Numerous cryptosystems operate algorithms that are utilized in the method of cryptography to preserve security. These systems are outlined on a group of specific parameters, known as a key, which is used in combination of original message as an input to the encrypting algorithm. The encrypted message together with the key serves as the input for the receiver when applying the decrypting algorithm. Fig. 3. Shows the process of cryptography.

Data security depends entirely on the secrecy of the key. Classically cryptography algorithms are divided into two forms depending on key distribution techniques:
  i. Symmetric Key Algorithms [6]
  ii. Asymmetric Key Algorithms [6]

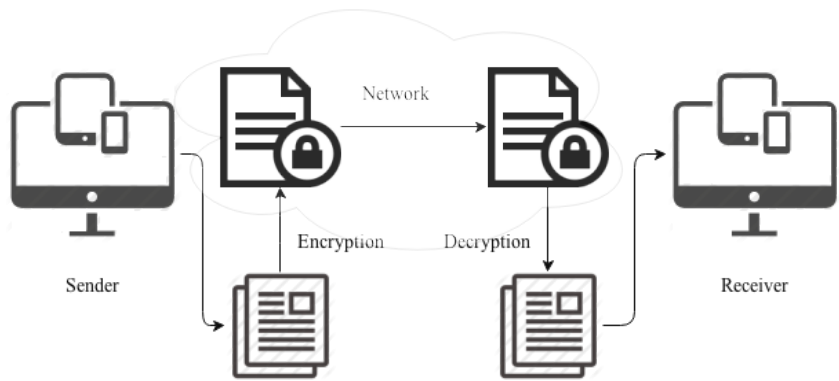

Figure 3. Process of Cryptography

Cryptanalysis on the other hand refers to study of cryptosystems with a view to finding weaknesses in them that will permit retrieval of the original message from the encrypted message, without the knowledge of the key or the algorithm used.

## 6.2. Classical Cryptography Techniques

### 1. Symmetric Key Cryptography

Cryptosystems that make use of symmetric key distribution use same key for encryption and decryption. This method is also known as secret key cryptography. Secure communication channel in key management is achieved only if the symmetric keys are pre-distributed in to every pair of interactive systems. Fig. 4. Shows the process of symmetric key cryptography.

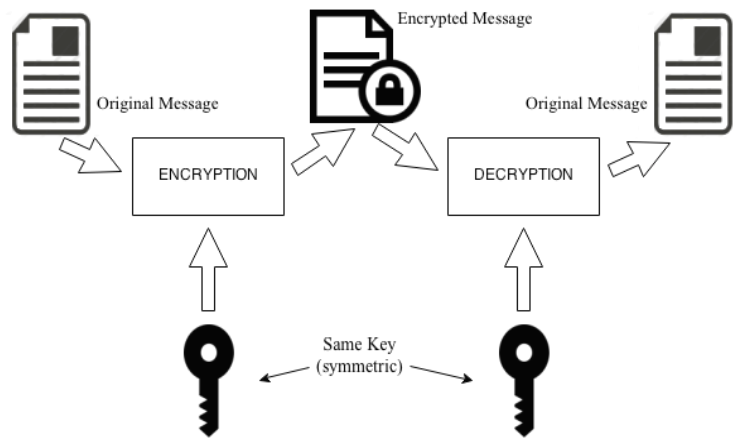

Figure 4. Symmetric Key Cryptography

### 2. Asymmetric Key Cryptography

Cryptosystems that make use of asymmetric key distribution use a public key system that consists of two parts: a Private key, which is kept secret and a Public key, which is distributed over the network. The sender encrypts the message using the public key of the receiver. The receiver makes use of its private key to decrypt the message. In such a distribution the private key is never in transit and hence less vulnerable to security issues. Fig. 5. Shows the process of asymmetric key cryptography.

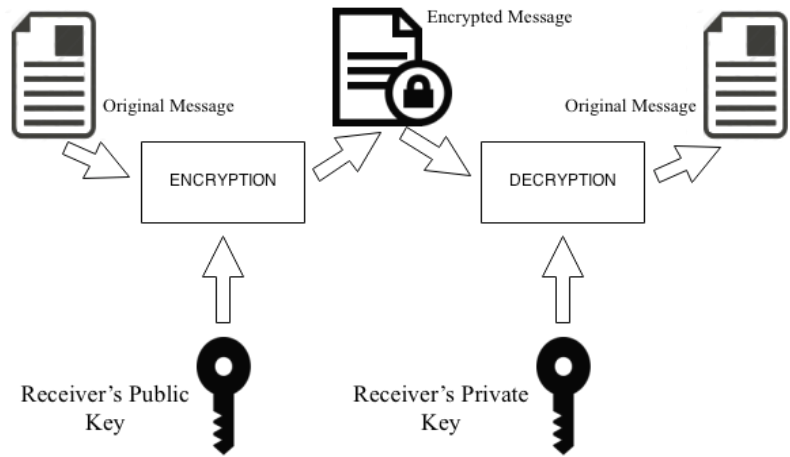

Figure 5. Asymmetric Key Cryptography

## 6.3. Limitations

Even though classical cryptography provide security in terms of privacy of message but this security is conditional. Some of the weaknesses that make this technique unsecure are as follows:

1) Authentication: Classical techniques do not provide a method to sender or receiver to find out the integrity of the message.

2) Easily hack-able: These techniques are based on computational complexity of the key, which can be cracked using mathematical algorithms (cryptanalysis).

3) Key distribution: The main weakness that this paper deals with is key distribution. There is no method in classical cryptosystems to discover the tampering with the key while transmitting enabling other users to read intercepted messages.

So to overcome the limitation of classical cryptosystems an efficient way of secure key distribution is proposed for communication by using quantum cryptography.

## 7. QUANTUM CRYPTOGRAPHY

### 7.1. Overview of Quantum Cryptography

Quantum cryptography [7] is an evolving technology that provides safety and security for network communication by performing cryptographic tasks using quantum mechanical effects. Quantum Key Distribution (QKD) is a technique that is an application of quantum cryptography that has gained popularity recently since it overcomes the flaws of conventional cryptography. QKD makes the secure distribution of the key among different parties possible by using properties of physics.

The quantum states of photons are used and the security key information is transmitted via polarized photons that contain the message denoted by bits (0 or 1) and each photon contains one bit of quantum information called as Qubit. The sender sends the polarized photon to the receiver. At the receiver end, the user determines the photon polarization by passing it through a filter and checks for any modifications in the received bits of photons when compared to the bits measured by the receiver. Any modifications found would show that there has been an intrusion from a third party because the intrusion would irreversibly change the encoded data in the photon of either the sender or the receiver. This method is based on the Heisenberg's uncertainty principle that states that the quantum state can't be measured without disturbing the state of either the sender or the receiver and hence introducing an anomaly in the quantum system that can be noticed by users as an intrusion.

Thus, Quantum cryptography applies the principles of physics governed by the laws of quantum mechanics for distributing the secret cryptographic key among the parties involved in the cryptosystem in a manner that makes it next to impossible for a third party to eavesdrop.

### 7.2. BB84 QKD Protocol

In order to facilitate QKD many protocols exist such as: BB84 [8], B92, Six-State, SARG04 [9], Ekert91. Among these protocols, BB84 is the most popular and widely used protocol for key distribution in practical systems.

Bennett and Brassard proposed BB84 protocol in 1984. The protocol consists of two main channels used for transmission:

1) Quantum channel: One-Way communication.
2) Classical channel: Two-way communication.

BB84 allows two parties conventionally a Sender and a Receiver to establish communication by a common key sequence using polarized photons.

Key exchange and key sifting are done as follows.

Using Quantum channel (Raw Key Exchange):

- The Sender encodes the information in random bits of 0 & 1 and uses randomly selected bases (rectilinear or diagonal) to transmit bits in it as shown in Fig. 6. Bases for each photon is chosen at random and the sender repeats this in order to send all the photons to the receiver.

- The receiver at the other end selects the basis (rectilinear or diagonal) at random to measure the received photons being unaware of the bases used to encode the photons by the sender.

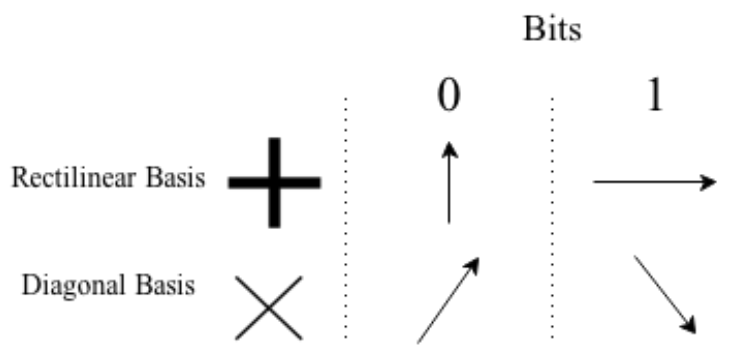

Figure 6. Photon Polarization using Bases

After receiving all the photons, the receiver now communicates with the sender using the public channel for key sifting.
Using Classical channel (Key Sifting):

- Receiver informs the Sender what bases he used to measure the photons and Sender responds by saying if it matched the bases used.

- Both agree on to the correct matching of the bases used and without announcing the actual value of information. After discarding all the data on the polarizer bases that did not match, both are left with two key strings of shorter sequences, known as the raw keys.

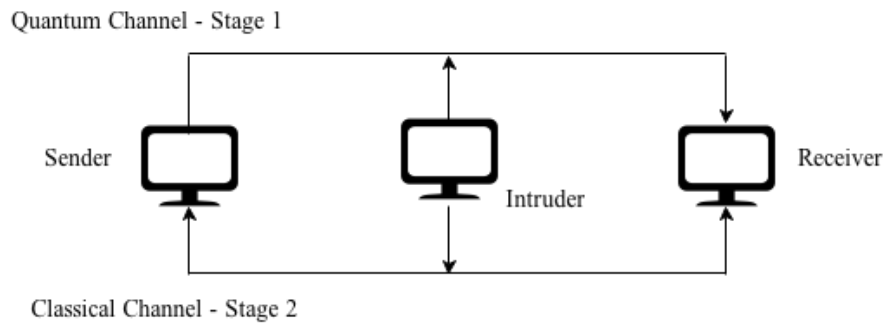

Figure 7. Quantum Key Distribution Setup

The bits that match make up the secret raw key which is not complete key and the communication still continues between the two comprising of the following steps:

*1) Error estimation:* In order to check if any eavesdropping has occurred, the raw keys are compared. If intrusion takes place, error would be introduced in one of the raw keys and the two keys on comparison won't match. Hence, the errors are to be estimated and if the error rate exceeds the threshold QBER (Quantum Bit Error Rate) for quantum transmission, the key is aborted and they try sending data again.

*2) Error correction (Reconciliation):* Performed to get the common key by removing the errors in the raw key by using a protocol from the many available protocols. The most widely used protocols are Cascade (based on optimal linear codes, uses and releases less data, end by performing parity based error correction), Winnow (based on exchange of parity and helps correcting single errors using hamming hash function).

*3) Privacy Amplification:* In the end both Sender and Receiver will hold, not completely private but identical strings of bits the information of which can be partially obtained on eavesdropping by a third party. The privacy amplification step helps remove this partially obtained information by the third party and hence make a correct secured secret key.

This paper applies an improved version of BB84 that is SARG04, which was proposed in 2004 by Scarani, Acin, Ribordy, and Gisin. SARG04 is an improved version of BB84 in terms of Photon-Number Splitting attacks. It uses laser pulses instead of single-photon sources used in BB84, which makes it robust protocol.

The first phase of this protocol remains similar to that of earlier mentioned BB84. The difference occurs in the second phase where the Receiver and the Sender compare their bit information to check for the matched basis and thus retrieval of secret key. In the second phase, unlike BB84, the Sender does not give the exact information of the basis used for encoding but instead provides the choice of two bases among which the Sender uses one of the bases. So the Receiver has to match with the correct bases used by the Sender to get measurement of the correct state and if it has used the incorrect bases, it won't be able to match any of the states of the Sender and hence wouldn't be able to access the data.

## 8. METHODOLOGY

### 8.1. IEEE 802.11 WLANs

The main objective of this paper is to offer secure key distribution in wireless networks making use of Quantum Cryptography. In order to properly facilitate the functioning of QKD it is found that IEEE 802.11 family best suits to be integrated with QKD. Fig. 8. Shows the architecture of IEEE 802.11.

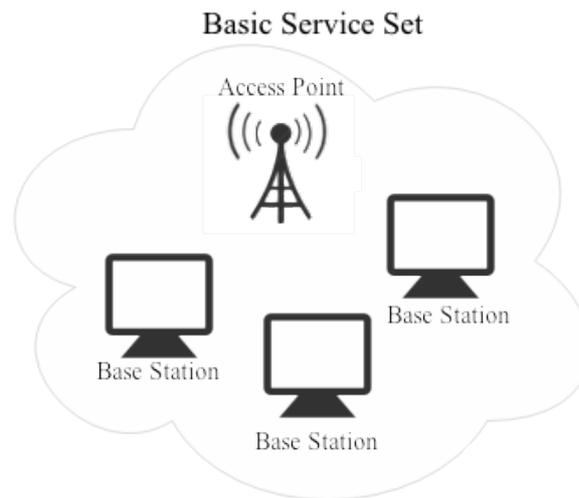

Figure 8. IEEE 802.11 Architecture (Single Cell)

Some of the characteristics of 802.11 WLANs that give it the proper environment for incorporating QKD are as follows:

1) Usage: 802.11 WLANs are mainly used in office and campus infrastructure, which facilitates the deployment of QKD network with a high density of quantum apparatus if necessary.
2) Capacity: Terminals present in a 802.11 WLAN have more computational capacity and more energy as compared to those in cellular networks.
3) Connection: 802.11 WLANs are used to provide access to the Internet through an AP installed in an organization, which is beneficial for QKD integration.

In order to facilitate efficient authentication and management of keys between access point and client, along with user traffic control 802.11 networks employs Extensible Authentication Protocol (EAP) [11]. EAP provides an authentication framework, which will be used in the current work. The security of 802.11 WLANs is based on the WEP protocol, with enhanced security being provided by the MAC layer. As specified earlier WEP presents security problems in a wireless network. The MAC layer on the other hand defines two classes for security: Robust Security Network Association (RSNA) [10] and Transaction Security Network (TSN) [10].

## 8.2. Robust Security Network Association

The RSNA defines two types of key hierarchies in order to divide the initial key material for protection of data frames. These two key hierarchies are:
    *i. Pair wise key:* To protect unicast traffic.
    *ii. Group wise key:* To protect multicast traffic.

This paper deals with Pair wise key hierarchy as a means to integrate QKD. Pair wise key hierarchy is represented in Fig. 9.

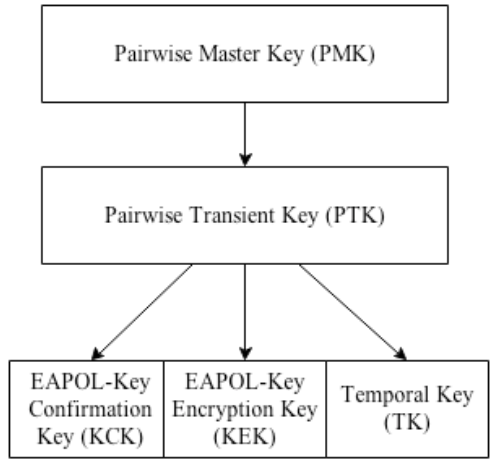

Figure 9. Pairwise Key Hierarchy

The PMK is received by the authentication server is the root key through which other transient keys are derived using pseudo-random function and must be protected. These transient keys are further divided into three keys.
    i. EAPOL-Key Confirmation Key (KCK): Used for data origin authenticity and calculates Message Integrity Code (MIC).
    ii. EAPOL-Key Encryption Key (KEK): Used for privacy and to encrypt Group Temporal Key (GTK)
    iii. Temporal Key (TK): Used in unicast data transfer for encryption.

In order to support the proper distribution and management of EAPOL-Keys RSNA defines a protocol called 4-Way Handshake.

## 8.3. 4-Way Handshaking

The 4-Way handshake performs the authentication process in IEEE 802.11 networks. The process allows the AP and the BS to generate the key hierarchy in order to provide encryption for secure communication. Since the keys are generated using a pseudorandom function, in order to further randomize data two random nonce values are transmitted between the AP (ANonce) and BS (SNonce). 4-Way handshaking process is shown in Fig. 10.

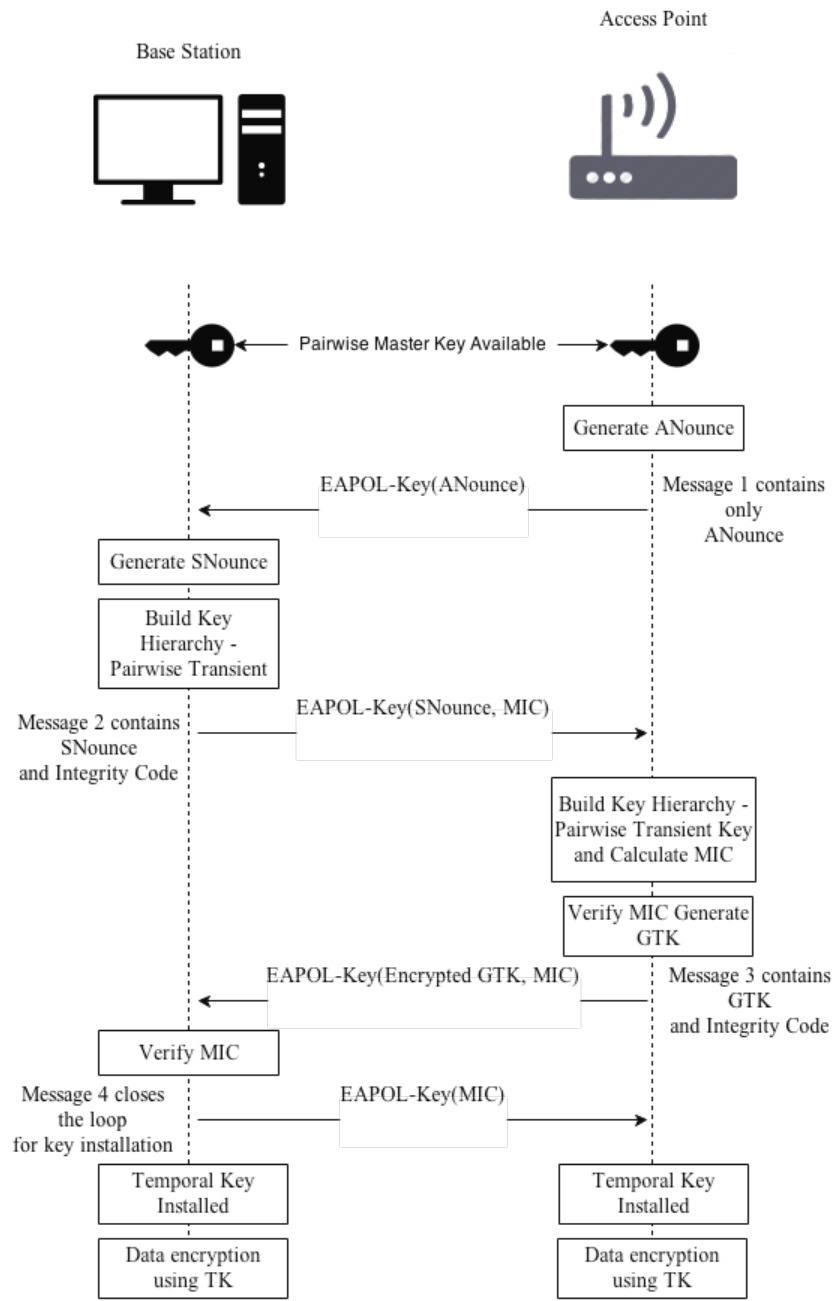

Figure 10. 4-Way Handshaking

4-way handshake itself has some flaws that inhibit secure communication in a network. Some of the vulnerabilities are [22-23]:
1) *Security Attacks:* Vulnerable to attacks like DoS in which intruders can torrent message to the BS after the handshake is completed.
2) *Response Time:* BS will disassociate and de-authenticate AP if data flow not received within the expected time interval after a successful handshake.
3) *Air cracking:* Keys can be recovered using this program once enough data packets have been captured.

In order to overcome these weaknesses we propose an integration of quantum cryptography in this key distribution mechanism as presented in the following subsection.

## 8.4. Proposed Protocol

In order to overcome the security issues of key distribution this paper employs SARG04 QKD protocol, which is an improved version of BB84 as discussed in Section 6. Fig. 11 shows the SARG04 QKD protocol being implemented in the proposed 4-way handshake protocol.

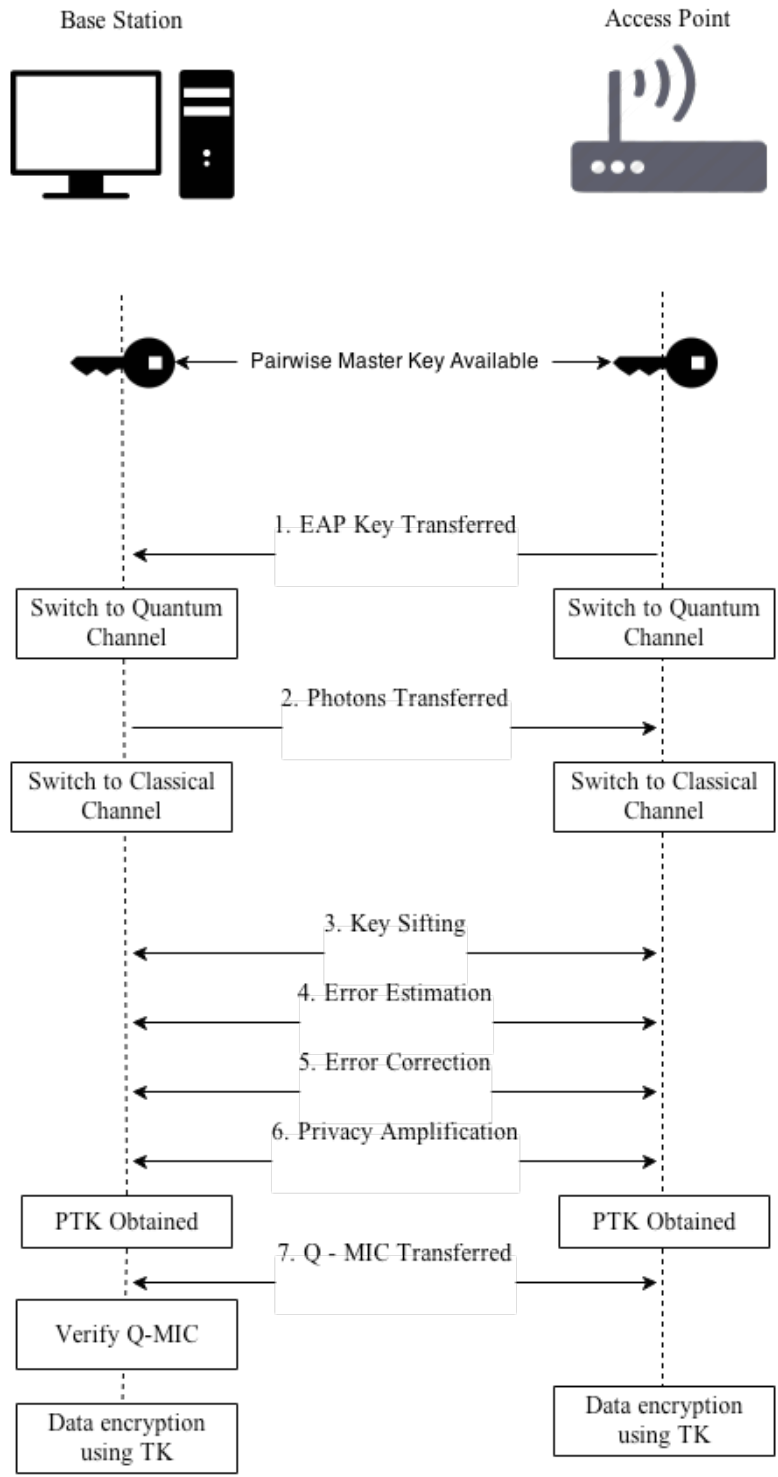

Figure 11. Proposed 4-Way Handshaking Protocol

The steps involved in the proposed protocol are as follows:

1. Initially the PMK is shared between the BS and the AP. Then the transmission process is switched to the quantum channel.
2. BS sends all the polarized photon to the AP using bases at random. As soon as the transmission of photons are finished, the channel is switched to classical.
3. The next 3 stages of QKD are applied to remove all the error and obtain the final encryption key.

In order to have secure communication, quantum transmission should make sure to send sufficient number of photons so as to improve the quantum key. PTK key (256 bits) is obtained by stripping the quantum key (384 bits) to match the number of bits. Once the PTK key is obtained the normal 4-way handshake protocol is followed wherein the MIC is derived for mutual authentication. The MIC is operated using an XOR operation and the first part of bits of equal length in PMK. This MIC is called as Quantum MIC (Q-MIC)

Q-MIC is transferred between both parties and verified. After which the temporal key is used to start encryption and communicate. Since the QKD protocol is used, Nounce values used in the original handshake are not required. Other advantages of applying this protocol are specified in the following section.

## 9. ADVANTAGES OF QUANTUM CRYPTOGRAPHY IN WIRELESS NETWORKS

Quantum cryptography is still a long way from being used in information transfer, since the real world implementation is far from the theoretical technique. But still there is no debate that quantum cryptography is a true breakthrough in network security. Some of the advantages of applying quantum cryptography in wireless networks are as follows [12]:

- Quantum Cryptography approach doesn't depend on mathematics models but instead is based on physics principles for decoding an encoded data making it virtually non-hack able and requires few resources for its maintenance.

- QKD is found useful for different categories of wireless networks as a means for connecting devices to different access points in close proximity. Hence making it more efficient for communication with fewer cables.

- Intrusion-eavesdropping can be detected or prevented from the vision of users only since the range under the coverage of wireless networks isn't too large in case of WPAN networks while for WLANs, the 802.11 wireless networks expand to the larger areas taking much benefit of the Quantum Cryptography.

Wi-Fi network usage is developing day by day also including the usage of hot-spot services very often. Since varied services (fields of national defense, aircraft communication etc.) are performed using Wi-Fi and hotspot its security is very important for all level of users and QKD has the ability of providing this security at the highest levels.

## 10. SPECIFIC CONTRIBUTION TO RESEARCH

The aim of the proposed solution is to provide enhanced security to the existing IEEE 802.11 wireless networks. In order to accomplish this a novel protocol has been developed to distribute secret key needed to encrypt data. A strong encryption key provides for an unconditional security throughout the communication session.

The modification to the existing IEEE 802.11i protocol has been done without impacting the existing frame format. The methodology is applicable for users who are equipped with quantum devices, but if they are not they can still continue with existing Wi-Fi communication.

The approach has vast amounts of future scope and possible extensions as well that are discussed in the following section. In essence, this research raises security to a new level with the absolute key security offered by Quantum Cryptography.

## 11. ALTERNATIVE APPROACHES

Besides standard and conventional methods of security services to users some of the other effective ways to secure WLANs are as follows [13]:

1) *Discovery and Mitigation of Rogue WLANs:* Unauthorized WLANs, which include APs and BS that are vulnerable, should be identified and must be controlled using an administrator in order to provide better security.

2) *Encryption and Authentication:* As specified throughout this paper encryption and authentication in WLAN are essential elements for WLAN security.

3) *Disabling Remote Logins:* This technique allows for safer secure means of communication without the concern of eavesdropping.

4) *Network Firewall and Intrusion Detection:* Traffic in WLANs should be safeguarded by means network firewall, network antivirus scanner or intrusion detection systems.

All these ways combined correctly provide a holistic approach to network security for wireless users.

## 12. CONCLUSION

The main goal of this research work is to show a method to improve the security aspect of WLANs. It has been shown that the integration of Quantum Cryptography in Wireless Networks has great prospective in terms of better network security.

Key management and distribution is difficult using classical cryptographic algorithms but the proposed approach provides a better solution for this problem. Research has shown that use of QKD to distribute network key raises the security and makes it harder for an eavesdropper to interrupt communication. With the proposed modification, this paper has achieved the main objective of improving security of WLANs.

**Authors**

**Ms. Priyanka Bhatia** is currently pursuing B.E (Hons) from Birla Institute of Technology & Science (BITS), Pilani in the department of Computer Science in Dubai, UAE

**Mr. Ronak Sumbaly** is currently pursuing B.E (Hons) from Birla Institute of Technology & Science (BITS), Pilani in the department of Computer Science in Dubai, UAE